\documentstyle[preprint,aps]{revtex}
\begin{document}
%\draft

\title{
{\bf COMMENT ON TRANSVERSE MASS DEPENDENCE OF PARTONIC DILEPTON PRODUCTION
IN ULTRA-RELATIVISTIC HEAVY ION COLLISIONS}
}

\author{{\bf K. Geiger} \\
{\it CERN TH-Division, CH-1211 Geneva 23, Switzerland}
}

%\begin{flushright}
%NSF-ITP-93-134
%\end{flushright}

\date{\today}

\maketitle

\begin{abstract}
Response to a recent Comment by M. Asakawa
concerning scale-breaking effects in dilepton emission
from partons in pre-equilibrium and quark-gluon plasma
during the early stage of ultra-relativistic heavy ion collisions.
\end{abstract}
\noindent

\vspace{2.5cm}

PACS Indices: 12.38 Mh, 25.75.+r, 24.85.+p, 13.87.Fh
\vspace{0.5cm}
\bigskip

In a recent paper M. Asakawa \cite{ref1} commented
a previous letter \cite{ref2} on $M_\perp$-dependence of
dilepton emission \cite{ref3} calculated with in the framework
of the Parton Cascade Model (PCM) \cite{ref4}.
He suggested that the $M_\perp$ scale breaking effects
discussed in \cite{ref2} may possibly be of unphysical
origin, arising from the perturbative QCD infrared cut-offs
($\mu_0$ and $p_{\perp cut}$) inherent
to the PCM.
I would like to respond first with a number of general remarks, followed
by some comments on the PCM approach in particular.
\begin{description}
\item[1.]
It is a fact that QCD is {\it not} scale invariant, even for massless
particles.
The characteristic scale is set by the "glueball mass" associated with the
gluon condensate, which can be interpreted phenomenologically e.g. in terms of
the string constant, the energy density residing in the gluon field, being
determined to be of the order of $\kappa \simeq 1$ GeV/fm$^2$.
Thus, in any perturbative QCD description that does not account for
the rather little understood non-perturbative mechanisms, one must inevitably
introduce some invariant mass cut-off around 1 GeV that separates the
perturbative regime from the non-perturbative domain. However, this is
not an arbitrary, unphysical parameter, but rather reflects the fact that there
is a fundamental scale in the problem. In particle physics phenomenology
this scale is usually associated with the string constant,
the hadronic radius or the typical hadronization time scale $\approx 1$ fm/c.
\item[2.]
In addition to the above natural QCD scale one is faced with further
(external) scale breaking quantities when addressing nucleus-nucleus collisions
that modify QCD processes in nuclear matter as compared to free space.
(i) The nuclear radii $R_A$ and the collision geometry define a finite size
system  that give rise volume and surface effects.
(ii) The nuclear density $\rho_A$ together with the Lorentz contraction
at high energies leads to a initial quark and gluon density already
in the initial state.
This in turn determines the initial condition for the
mean free path or the mean collision frequency of parton interactions.
(iii) When following a self-contained
dynamical description of nuclear collisions, the materialization and
excitation of partons and their energy dissipation will clearly
increase the initial density, leading at sufficiantly large
beam energy to a hot and very dense parton matter.
Thus one has a time-dependent density and temperature as further
scales in the problem both of
which are externally determined by the collision energy, and nuclear size.
\item[3.]
In a description of high energy nuclear reactions
on the basis of "scaling hydrodynamics" one assumes a priory an ideal
fluid dynamical expansion of the matter produced in the central
collision region. That is, one assumes local thermalization, a
longitudinally boost invariant expansion, absence of radial flow, and
no other scales than the temperature are involved in the dynamical
evolution \cite{mcl}.
In other words, all of the above mentioned scale breaking quantities
are completely ignored here -  an approach which is certainly not illegitimate,
but should not be taken as a measure of realistic description.
It is interesting to note that a recent study by M. Strickland
\cite{ms} of the dependence on
initial conditions for thermal dilepton production concealed that
even the later stages of the thermal expansion will carry
information about the initial conditions. Thus, even a presiumed
thermal evolution must account for tose external parameters that
determine the initial state and the necessarily generate in scale
breaking effects.
\end{description}

\noindent
Responding to the specific points of Asakawas Comment, I state the
following:
\begin{description}
\item[4.]
{\it Mass cut-off $\mu_0$}:
The argument that by a time of about $1/\mu_0 = 0.2$ fm/c the partons
have reached this invariant mass cut-off and propagate on as massive
($\approx 1$ GeV) particles is definitely not correct,
because this estimate is based
on "free" cascading of virtual partons by successive bremsstrahlung,
e.g. in jet evolution of $e^+e^-$ annihilation. In a nuclear collision,
that is in the dense matter environment of the central collision region, this
gradual deexcitation
of virtual particles toward mass shell is considerably delayed
due to scatterings and fusions. This has been recently addressed
in a different context \cite{ref5}
and is currently investigated in more detail.
The more frequent these interactions with the environment is,
i.e. the denser the matter is packed, the longer it takes for a parton
to reach $\mu_0$. In the PCM calculations for Au+Au at RHIC, it takes about
1 fm/c (!) until the majority of materialized virtual partons do
not radiate anymore because they have reached $\mu_0$.
I strongly object the summarizing remark in \cite{ref1} stating
that the PCM relies on "...the use of
inappropriately large vitualities $\ge \mu_0$...". It is well known that
in free space jet evolution, the mass $\mu_0$ of the order of a GeV
is the typical scale at which non-perturbative
collective effects begin to establish a pre-confinement \cite{basetto}.
On the other hand in a dense matter environment, the screening effects
generate a Debye mass that corresponds to an effective mass of a particle in
the medium, and which takes over the role of $\mu_0$.
Of course, in the PCM the parameter $\mu_0$ and also
the minimum allowed momentum transfer $p_{\perp cut}$ in parton collisions
are partly responsible for the $M_\perp$ scale breaking in the dilepton
spectrum.
This has been clearly stated in Ref. \cite{ref2} on p. 3078.
However, this scale breaking contribution is a physical effect
and is intimately connected to the
(medium dependent) Sudakov formfactors of the partons that are a
direct consequence of
the renormalization group improved parton picture. Therefore
- in accord with item 1. - it is a physical manifestation of the
fundamental fact that QCD is not a scale invariant theory.
\item[5.]
{\it $q\bar q$ annihilation}:
As stated in Ref. \cite{ref3} on p. 1922, the turnover
at lower dilepton mass is due to the neglect of contributions
from quark antiquark scatterings which are treated with the
phenomenological scattering amplitude, if the momentum transfer of
a parton collision is below $p_{\perp cut}$. These
low $p_\perp$ processes were not included in the calculation of
the dilepton spectrum, because perturbative QCD does not tell us
about the soft physics of these contributions, and they
are therefore model dependent. In order not to spoil the results for
the perturbative QCD yield of dileptons
where the amplitudes are well known,
a phenomenological description of soft production was avoided.
Notwithstanding the fact of
uncertainty in how to calculate the radiation from those softer
processes, I am sure that the dilepton mass spectrum would continue
to rise at low masses.
I doubt that the quark virtualities have much to do with it.
On the other hand, Asakawa is ultimately correct in saying that the PCM
predictions for dilepton masses less than about 3 GeV should not be
taken seriously at this time. But this was stressed in the paper too.
\end{description}

In conclusion,
I think that one has to be careful when comparing the PCM
calculations with e.g. the solutions of the Bjorken hydrodynamical
model. The latter  cannot account for the scale breaking effects
discussed above, because it assumes a priori that those are absent.
On the other hand, I am fully aware that the
PCM results are plagued by a number of uncertainties
which have been repeatedly discussed in preceding papers. The PCM
is not to be misunderstood as a fine tuned 'event generator'. Rather
than that its main purpose - at least at the current stage - is to approach
in an explorative way the complicated many particle aspects of
quark and gluon dynamics in heavy ion collisions. From a truly
quantitative picture one is still far away.


\bigskip

\begin{thebibliography}{5}

\bibitem{ref1}
M. Asakawa, {\it Comment on 'Transverse Mass Dependence of
Dilepton Emission from Pre-equilibrium and Quark Gluon Plasma
in High Energy Nucleus-Nucleus Collisions'}, LBL-35269 (1994),
also on hep bulletin board, nucl-th/9402033.

\bibitem{ref2}
K. Geiger, Phys. Rev. Lett. {\bf 71}, 3075 (1993).

\bibitem{ref3}
K. Geiger and J. I. Kapusta, Phys. Rev. Lett. {\bf 70}, 1920
(1993).

\bibitem{ref4}
K. Geiger and B. M\"uller,  Nucl. Phys. {\bf B369}, 600 (1992);
K. Geiger, Phys. Rev. D {\bf 47}, 133 (1993).


\bibitem{mcl}
L. McLerran and T. Toimela, Phys. Rev. {\bf D31}, 545 (1985).

\bibitem{ms}
M. T. Strickland, {\it Quark-Gluon Plasma Initial Conditions
from Thermal Gamma and Dilepton Rates}, preprint DUKE-TH-93-57
(to appear in phys. Rev. C).

\bibitem{ref5}
K. Geiger and B. Mueller, {\it QCD evolution equations for high energy
partons in nuclear matter}, preprints NSFITP-93-149, CERN-TH-7205/94;
K. Geiger, {\it Analytical solutions of QCD evolution equations
for parton cascades inside nuclear matter},
preprint CERN-TH-7206/94.

\bibitem{basetto}
See e.g.: A. Basetto, M. Ciafaloni and G. Marchesini,
Phys. Rep. {\bf 100}, 201 (1983).

\end{thebibliography}
\end{document}